\documentclass[journal,twoside,web]{ieeecolor}
\usepackage{tmi}
\usepackage{cite}
\usepackage{amsmath,amssymb,amsfonts}
\usepackage{graphicx}
\usepackage{textcomp}
\usepackage[ruled]{algorithm2e}
\usepackage{multirow}

\def\BibTeX{{\rm B\kern-.05em{\sc i\kern-.025em b}\kern-.08em
    T\kern-.1667em\lower.7ex\hbox{E}\kern-.125emX}}
\begin{document}
\title{Self-Supervised Federated Learning for \\Fast MR Imaging}
\author{Juan Zou, Cheng Li, Ruoyou Wu, Tingrui Pei, Hairong Zheng and Shanshan Wang
\thanks{Manuscript Prepared in March 2023; Submitted in May 2023; Corresponding author: Shanshan Wang.\\
Juan Zou is with the School of Physics and Optoelectronics, Xiangtan University, Xiangtan 411105, China, and the Paul C. Lauterbur Research Center for Biomedical Imaging, Shenzhen Institute of Advanced Technology, Chinese Academy of Sciences, Shenzhen 518055, China. (e-mail:zjuan@xtu.edu.cn).}
\thanks{Cheng Li, Ruoyou Wu, Hairong Zheng and Shanshan Wang are with the Paul C. Lauterbur Research Center for Biomedical Imaging, Shenzhen Institute of Advanced Technology, Chinese Academy of Sciences, Shenzhen 518055, China. (e-mail: cheng.li6@siat.ac.cn;  ry.wu@siat.ac.cn;
hr.zheng@siat.ac.cn; ss.wang@siat.ac.cn).
}
\thanks{Tingrui Pei is with the College of Information Science and Technology, Jinan University, Guangzhou 510631, China, and the School of Physics and Optoelectronics, Xiangtan University, Xiangtan 411105, China. (e-mail:peitr@163.com).
}
}
\maketitle

\begin{abstract}
Federated learning (FL) based magnetic resonance (MR) image reconstruction can facilitate learning valuable priors from multi-site institutions without violating patient’s privacy for accelerating MR imaging. However, existing methods rely on fully sampled data for collaborative training of the model. The client that only possesses undersampled data can neither participate in FL nor benefit from other clients. Furthermore, heterogeneous data distributions hinder FL from training an effective deep learning reconstruction model and thus cause performance degradation. To address these issues, we propose a Self-Supervised Federated Learning method (SSFedMRI). SSFedMRI explores the physics-based contrastive reconstruction networks in each client to realize cross-site collaborative training in the absence of fully sampled data. Furthermore, a personalized soft update scheme is designed to simultaneously capture the global shared representations among different centers and maintain the specific data distribution of each client. The proposed method is evaluated on four datasets and compared to the latest state-of-the-art approaches. Experimental results demonstrate that SSFedMRI possesses strong capability in reconstructing accurate MR images both visually and quantitatively on both in-distribution and out-of-distribution datasets.

\end{abstract}

\begin{IEEEkeywords}
MRI reconstruction, federated learning, self-supervised learning.
\end{IEEEkeywords}

\section{Introduction}
\label{sec:introduction}
\IEEEPARstart{M}{agnetic} resonance imaging (MRI) is an effective tool in the clinic for disease diagnosis, treatment planning, and prognosis evaluation. However, since the MRI data acquisition process needs to traverse a sequence of paths in k-space, the imaging speed is slow, which has been a bottleneck limiting factor for the wide adoption of MRI in routine clinical practice. Undersampled data acquisition is an effective technique to accelerate MR imaging \cite{chartrand2009fast,zhao2015accelerated,lustig2007sparse}. To reconstruct high-quality MR images from undersampled k-space data, deep learning methods have been widely used \cite{wang2016accelerating}. The performance of deep learning models depends heavily on the dataset utilized. Large quantities of heterogeneous paired undersampled and fully sampled data are required to properly train a deep learning model, whereas collecting these data at a single institution is challenging. Multi-institutional collaboration is necessary in this case.

One method to utilize multi-institutional data is gathering the data together and training a centralized model on this large dataset. Although the method is straightforward and effective, it involves a data privacy issue. Federated learning (FL) methods are proposed to specifically address this issue \cite{yang2019federated,li2019privacy}. Unlike centralized training, FL methods utilize decentralized data in multiple institutions to collaboratively learn a global model without sharing any local raw data. One classic FL method, FedAvg, obtains the weights of the global model by averaging the weights of local models during each communication round \cite{mcmahan2017communication}. Although this type of FL can exploit multi-institutional data while protecting data privacy, they suffer from a notable data heterogeneity issue. The multi-institutional data are probably acquired with different imaging protocols. Large discrepancies may exist between the data distributions, which can hinder the learning of an effective global model and in turn, lead to inferior performance for each client \cite{shi2022uncertainty}.

Recently, several methods have been proposed to address the issue of data heterogeneity in FL for accelerating MR imaging \cite{guo2021multi,feng2022specificity,elmas2022federated}. In \cite{guo2021multi}, adversarial alignment between source and target sites was proposed to improve the similarity of latent-space representations in reconstruction models. In \cite{feng2022specificity} and \cite{elmas2022federated}, the reconstruction models (unconditional GAN and U-Net) were divided into an aggregated part and an unaggregated part to learn the generalized and client-specific representations. These methods have achieved better reconstruction performances than traditional federated learning methods. However, all these methods adopted the fully-supervised training approach, and they assumed that enough fully sampled data exist in each institution. In real situations, according to the setting of normal scanning protocols, the acquisition of fully sample data in hospitals is rare \cite{millard2022framework}. Therefore, developing FL methods that only utilize undersampled data is desired for real clinical applications.

To address the aforementioned challenges, we propose a self-supervised federated learning method for accelerating MR imaging (SSFedMRI). SSFedMRI can solve the issues of fully sampled data deficiency on each client, achieving cross-site collaborative training with only undersampled data. Specifically, we re-undersample the acquired undersampled k-space data on each client, set them to physics-based contrastive reconstruction networks, and perform local training with designed self-supervised loss. Then, the updated local model will be aggregated into a global model in each communication round. Because the employed physics-based reconstruction networks have fewer parameters than existing FL-based methods, which apply data-based reconstruction network as base framework in clients, reduced communication costs are expected. Heterogeneous data distribution usually causes performance degradation in FL. To address this issue, we propose a personalized soft update scheme. Specifically, the main contributions of this paper can be summarized as follows:

\begin{itemize}
    \item A self-supervised FL framework for MRI reconstruction is proposed (SSFedMRI), which leverages undersampled data from multiple institutions to learn prior knowledge without relying on fully sampled reference data. SSFedMRI mitigates insufficient undersampled data in local client and preserves data privacy. In addition, thanks to the  physics based contrastive learning framework \cite{wang2022parcel}, the communication costs have been reduced compared to pure data-driven FL frameworks.
    \item A personalized soft update scheme is designed to simultaneously capture the global shared representations among different centers and maintain the specific data distribution of each client. Instead of directly replacing the weights of local models with those of the global model during each communication round, we update the local models according to the discrepancies between the global model and the local models. In this way, the local models can learn cross-site prior information from participant clients while still holding their optimization towards local data distribution.
    \item Comprehensive experiments on multi-site MR complex-valued datasets have been performed. Experimental results demonstrate that our proposed method, SSFedMRI can achieve better performances both visually and quantitatively compared to state-of-the-art approaches. Moreover, the communication cost of SSFedMRI is also much lower than those of existing methods.
\end{itemize}

\section{Related Work}
\subsection{Deep MRI reconstruction}
 Deep MRI reconstruction learns prior knowledge from big datasets to solve the ill-posed inverse problem of reconstructing MR images from undersampled k-space data. Lots of existing reviews categorize existing deep MRI reconstruction approaches into data-based methods and physics-based methods \cite{liang2020deep,wang2021deep}. Data-based methods rely on deep neural networks and big datasets to learn the mapping between undersampled data and fully sampled data \cite{eo2018kiki,wang2020deepcomplexmri,mardani2018deep}. These models need to be re-trained each time when new data with different distributions are acquired. Physics-based methods, such as VN \cite{hammernik2018learning}, ADMM \cite{sun2016deep}  MoDL \cite{aggarwal2018modl} and ISTA \cite{zhang2018ista}, construct network architectures based on respective iterative optimization algorithms, which can incorporate the physics of MRI. These methods have achieved outstanding performances under the supervised learning paradigm. 

In the meantime, some recent studies have tried to address the issue of fully sampled data deficiency. Various self-supervised methods have been proposed by directly treating the undersampled raw data or variants as the supervisory signals 
 \cite{yaman2020self,wang2022parcel}, including SSDU \cite{yaman2020self} and PARCEL \cite{wang2022parcel}. Nevertheless, heterogeneous data are needed to train a robust deep learning model, and it is difficult if not impossible to acquire enough data with rich diversities in a single institution. Multi-institutional collaboration without violating the patients’ privacies is a megatrend.

\subsection{Federated learning}
Without gathering data together, FL exploits multi-institutional prior information by learning a global model while aggregating locally trained models. The standard FL method aggregates local models by averaging the model weights in each round repeatedly until the global model converges. Domain shift is a common challenge encountered in FL, which is caused by the data heterogeneity among clients. To solve this challenge, various personalized FL approaches have been proposed \cite{li2021fedphp,xu2022closing,feng2022specificity,elmas2022federated,collins2021exploiting,fallah2020personalized,levac2023federated} (e.g., methods involving model interpolation \cite{li2021fedphp,xu2022closing}, network split \cite{feng2022specificity,elmas2022federated,collins2021exploiting}, and local fine-tuning \cite{fallah2020personalized,levac2023federated}), targeting to learn personalized local models according to the specific data distributions.

Protecting patients’ privacies is crucial in medical imaging, and thus, building FL models are essential. FL has been adopted in different medical imaging tasks  \cite{kaissis2021end,bercea2022federated,dayan2021federated,liu2021feddg,guo2021multi,feng2022specificity,elmas2022federated,levac2023federated}, including classification \cite{kaissis2021end,bercea2022federated,dayan2021federated}, segmentation \cite{liu2021feddg}, and reconstruction \cite{guo2021multi,feng2022specificity,elmas2022federated,levac2023federated}. For MRI reconstruction, Guo et al. \cite{guo2021multi} proposed adversarial alignment between source and target sites to reduce domain shift. Feng et al. \cite{feng2022specificity} and Elmas et al. \cite{elmas2022federated} splitted the reconstruction models (unconditional GAN and U-Net) into two parts to respectively learn the general and client-specific representations. Brett et al. \cite{levac2023federated} personalized client-sided physics-based reconstruction network in FL via fine tuning. These FL methods for MRI reconstruction have obtained promising reconstruction results. However, supervised learning was utilized by these methods, and each participating institution will need to acquire enough fully sampled reference data. Thus, there is an urgent need for the development of un-supervised FL methods for MRI reconstruction.

\subsection{Self-supervised federated learning}
Self-supervised federated learning methods learn representative information from cross-site unlabeled data with different training schemes. Among them, federated contrastive learning (FCL) is a popular direction, which combines contrastive learning with FL \cite{zhang2020federated,zhuang2021collaborative,wu2022distributed,dong2021federated,yan2023label}. For example, FedCA trained an alignment model to constrain the local model with shared dataset among clients \cite{zhang2020federated}. FedU choosed to update local models or not according to the divergence between local model of each client and global model \cite{zhuang2021collaborative}. Meanwhile, some FCL methods for medical imaging tasks have been proposed \cite{wu2022distributed,dong2021federated,yan2023label}. For example, Wu et.al. pre-trained a main encoder with unlabeled data by exchanging features among clients and fine-tune it with a small amount of labeled data for image segmentation \cite{wu2022distributed}. Dong et al. transferred metadata of local encoder feature vectors to other clients, and their framework can be fine-tuned for different tasks (detection, localization, and segmentation) locally in either supervised or semi-supervised fashion \cite{dong2021federated}. Yan et al. proposed a Transformer-based self-supervised FL framework via masked image for image classification \cite{yan2023label}. These studies are built for image post-processing analysis and cannot be directly extended to MRI reconstruction.

In this work, we propose a self-supervised federated learning method, SSFedMRI, for accelerating MR imaging.  Different from existing FCL-like works, our SSFedMRI neither needs any fully sampled reference data to fine-tune the pre-trained models nor shares any feature or raw data among clients. Furthemore, we softly update the local denoiser models with an “update” item to construct personalized denoiser models. This update item is determined by the discrepancies between the global model and the local models. To the best of our knowledge, we make the first attempt to address the issue of fully sampled data deficiency in FL for MRI reconstruction.

\section{Method}
\subsection{MR image reconstruction}

MRI reconstruction aims to recover an MR image 
$\mathrm{x} \in \mathbb{C}^N
$ from the undersampled k-space data $\mathrm{y} \in \mathbb{C}^M(M \ll N)$. The imaging scheme can be expressed as follow:
\begin{equation}\mathrm{y}=A  \mathrm{x}+\mathrm{e},\label{1}\end{equation}
where $\mathrm{e} \in \mathbb{C}^M
$ refers to the noise and $A=PF$ is an undersampled Fourier encoding operator. Recovering $\mathrm{x}$ from $\mathrm{y}$ is an ill-posed inverse problem. In general, image reconstruction is formulated as a regularized optimization problem:

\begin{equation}\mathrm{x}^{r e c}=\arg \min _{\mathrm{x}}\|A x-\mathrm{y}\|_2^2+\lambda 
 \mathcal{R}(\mathrm{x}).\label{2}\end{equation}
Here, $\mathcal{R}\left(\mathrm{x}\right)$ represents a regularization prior, which restrict the solutions to the space of desirable images. In physics-based deep learning methods, such as MoDL [35], Eq.\eqref{2} is solved iteratively via alternating minimization over $\mathrm{z}$ and $\mathrm{x}$:

\begin{equation}z_n=D\left(x_n\right),\label{3a}\end{equation}
\begin{equation}x_{n+1}=\arg \min _{\mathrm{x}}\|A x-\mathrm{y}\|_2^2+\lambda\left\|x-z_n\right\|^2,\label{3b}\end{equation}
where $D$ is a denoiser of $\mathrm{x}$. $D$ learns a CNN-based prior to remove alias artifacts and noise. $\mathrm{n}$ refers to the iteration of an unrolled network. $z_n$ is the intermediate reconstructed image at iteration $\mathrm{n}$. $\mathrm{x}_n$ is the reconstructed image at each iteration $\mathrm{n}$. Eq.\eqref{3b} is a data-consistency item. After training, the trained denoiser model $D_\Theta$ in Eq.\eqref{3a} will be retained for online MR reconstruction.

\subsection{Federated learning formulation}

FL further expands the space of solution via communication between global imaging model $D_{\Theta_g}$ and local imaging models $D_{\Theta_k}$ in multi-clients. Each communication round in FL consists of four steps: local model training, local model upload, global model aggregation and local model update. Local model training is performed by minimizing the local reconstruction loss:

\begin{equation}\mathcal{L}_k^{r e c}\left(S_k, A_k\right)=\frac{1}{K} \sum_{k=1}^K\left\|D_{\Theta_k}\left(A_k^{\prime} y_k\right)-x_k^{r e f}\right\|_2^2,\label{4}\end{equation}
where $\left(y_k, x_k^{r e f}\right)$
is the undersampled k-space data $\left(y_k\right)$ and corresponding fully sampled reference image $\left(x_k^{r e f}\right)$ in decentralized datasets $S_1,S_2,\ldots, S_k,\ldots,S_K$ ($k$ refers to the $k^{th}$ client). $A_k^{'}$ is the adjoint of $A_k$ that transform undersampled measurement $\mathrm{y}_k$ to the zero-filled reconstruction. After a certain number of local training epochs in each communication round, local models are uploaded to the server, and aggregated by averaging their local parameters to obtain the global model:

\begin{equation}\Theta_g=\frac{1}{K} \sum_{k=1}^K\Theta_k,\label{5}\end{equation}
where $\Theta_k$ is the parameters of the local model in client $k$. Typically, after the global model converges, it can then be used for online undersampled MRI reconstruction.

Federated learning learns the regularization prior $\mathcal{R}\left(\mathrm{x}\right)$ in Eq.\eqref{2} from datasets of different clients without gathering the distributed datasets, which preserves patients’ privacies. However, the above workflow of FL for MRI reconstruction has two issues: 1) Whether the client without fully sampled data can participate in FL and benefit from distributed datasets of other clients? 2) For each client, whether it is the best to apply the global model to overwrite entire parameters of local model in each communication round? Our self-supervised FL method for MRI reconstruction in this work is proposed to specifically address these concerns.

\subsection{The SSFedMRI framework}

\begin{figure*}[htbp] 
\centering
\includegraphics[width=0.7\textwidth]{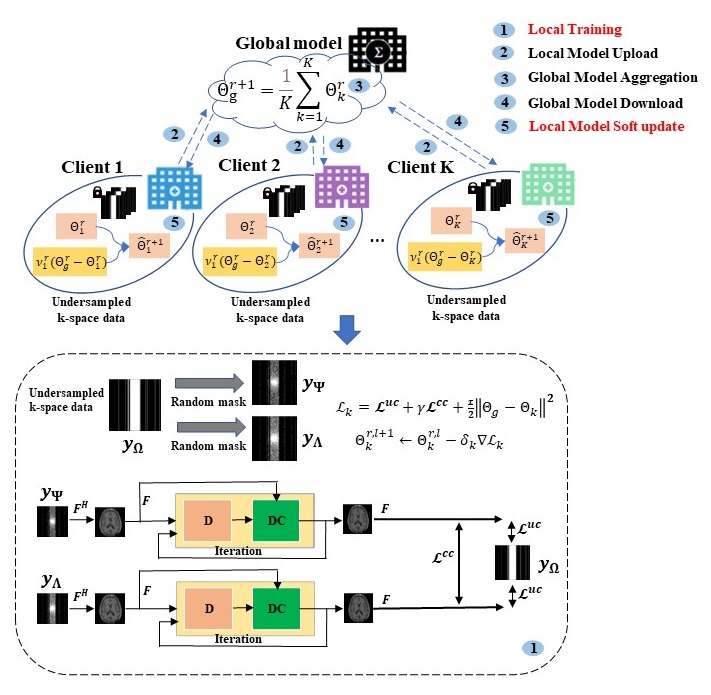} 
\caption{The framework of our proposed SSFedMRI. The local models in our self-supervised federated learning framework are optimized with three loss items, the undersampled consistency loss $L^{u c}$, the contrastive consistency loss $L^{c c}$, and a regulation term. $\mathrm{y}_{\Psi}$ and $\mathrm{y}_{\Lambda}$ are re-undersampled from the original raw k-space data $\mathrm{y}_{\Omega}$. Each reconstruction model in contrastive networks contains T iterations of the basic building block (a CNN-based denoiser and a data consistency layer). A soft update method is designed to update the local models during each communication round to ensure that denoiser priors from all clients are learned and in the meantime, the major gradient descent directions of local models are kept to properly fit local data distribution.}
\label{fig1}
\end{figure*}

The overall framework of our proposed SSFedMRI is shown in Fig. \ref{fig1}. SSFedMRI is constructed to tackle the issue of fully sampled data deficiency in FL for MRI reconstruction. To mitigate performance degradation due to data heterogeneity, we design a soft update in SSFedMRI to get the personalized model. SSFedMRI trains a global model with undersampled data supervision and personalized local models via soft update. The pipeline in each communication round of SSFedMRI consists of five steps: (1) local training – On each local client, physics-based contrastive reconstruction networks are constructed for MRI reconstruction. The acquired raw k-space data are re-undersampled into two subsets, which are treated as the inputs of the physics-based contrastive networks. The physics-based contrastive reconstruction networks are trained for $L$ epochs with an undersampled consistency loss, a contrastive consistency loss, and a regulation term (Eq. 6 and Eq. 7). (2) Model upload – After $L$ local training epochs, the parameters of the parallel reconstruction networks on each client are uploaded to the central server. (3) Model aggregation – The uploaded local contrastive
reconstruction networks are aggregated via FedAvg. (4) Global model downloaded – The aggregated global model is downloaded to each local client. (5) Local model update (soft update) – In each communication round, the local models are softly updated by adding corresponding update items, which are determined by the discrepancies between the global model and the local models. The details of the optimization process are provided in Algorithm 1.

\begin{algorithm}[]
	\caption{SSFedMRI}
	\LinesNumbered
	\KwIn{Undersampled datasets from $K$ clients: $S=\left\{S_1, \ldots, S_K\right\}$, number of communication rounds $R$, number of local epochs $L$, hyper-parameter $\mu$ and $\tau$ for loss in Eq. \eqref{6} and \eqref{7}, learning rate for local training $\delta_k$, hyper-parameter $\beta$.}
	\KwOut{Global model $\Theta_g$ and personalized models $\Theta_1,\dots,\Theta_k$.}
	Server sends $\Theta_g^0$ to all clients to initialize local models\;
	\For{$r=1,2,… ,R$}{
 \For{client $k \in S$ in parallel}{
		clients receive global model from server\;
  \eIf{$r=1$}{
			$\sigma_k \longleftarrow \frac{\beta}{\left\|\Theta_g^r-\Theta_k^r\right\|}$\;
   // overwrite local models with received global model\;$\widehat{\Theta}_k^{r+1} \longleftarrow \Theta_g^r$;
		}{
			$v_k^r \longleftarrow 1-\min \left(1, \sigma_k\left\|\Theta_g^r-\Theta_k^r\right\|\right)$\;
   // softly update local models to initialize personalized model;
   $\widehat{\Theta}_k^{r+1} \longleftarrow \Theta_k^r+v_k^r\left(\Theta_g^r-\Theta_k^r\right)$\;
		} 
  // conduct self-supervised loss for local training in Eq. \eqref{7};
  $\mathcal{L}_k=\mathcal{L}^{r e c}+\frac{\tau}{2}\left\|\Theta_g-\Theta_k\right\|^2$\;
  // train local models with above loss $\mathcal{L}_k$;
   $\Theta_k^{r+1} \longleftarrow \widehat{\Theta}_k^{r+1}- \eta_k \nabla \mathcal{L}_k$\;
  }
// average local models to acquire global model;
$\Theta_g^{r+1}=\frac{1}{K} \sum_{k=1}^K \Theta_k^{r+1}$\;
}
\end{algorithm}

\subsubsection{Local training}

Previous FL methods for MRI reconstruction performed supervised learning in the local training step, requiring fully sampled reference data on each client. Here, in SSFedMRI, we perform self-supervised learning with physics-based contrastive reconstruction networks instead. Two different sets of re-undersampled data are extracted from the raw undersampled k-space data, which are treated as the inputs to the contrastive networks. The physics-based contrastive reconstruction networks consist of two same models, which have the same network architecture but with different parameters.

The corresponding zero-filled images of the re-undersampled data are used as the inputs to the physics-based contrastive reconstruction networks. The role of the contrastive networks is to constrain the solution space by enforcing a contrastive consistency loss $L^{c c}$ between the two outputs. In addition to $L^{c c}$, we include an undersampled consistency loss $L^{u c}$ in our local training to ensure that the networks can learn essential and inherent representations in MR images. These two losses are combined to form the following reconstruction loss:

\begin{equation}\mathcal{L}^{\text {rec }}=\underbrace{\left\|\mathrm{y}_{\Psi \rightarrow \Omega}-\mathrm{y}_{\Omega}\right\|_2^2+\left\|\mathrm{y}_{\Lambda \rightarrow \Omega}-\mathrm{y}_{\Omega}\right\|_2^2}_{\mathcal{L}^{u c}}+\gamma \underbrace{\left\|\bar{\mathrm{y}}_{\Psi \rightarrow \Omega}-\bar{\mathrm{y}}_{\Lambda \rightarrow \Omega}\right\|_2^2}_{\mathcal{L}^{c c}},\label{6}\end{equation}
where $\gamma$ is applied to balance the contributions of undersampled consistency loss and contrastive consistency loss. $\mathrm{y}_{\Omega}$ refers to the acquired raw undersampled k-space data. $\mathrm{y}_{\Psi \rightarrow \Omega}$ and $\mathrm{y}_{\Lambda \rightarrow \Omega}$ are predictions of the two networks on sampled data points in $\mathrm{y}_{\Omega}$ (during the re-undersampling process), respectively. $\bar{\mathrm{y}}_{\Psi \rightarrow \Omega}$ and $\bar{\mathrm{y}}_{\Lambda \rightarrow \Omega}$ are the predictions corresponding to the unsampled points in $\mathrm{y}_{\Omega}$.

Besides, a regularization term, which is proportional to the $L_2$ norm of the discrepancy between the local model and the global model, is added to constrain the learning of each local model. This regularization term can guide the optimization of local models towards learning global prior information instead of focusing only on the local datasets during the local training period, preventing the offset between $\Theta_g$ and $\Theta_k$ to continuously increase during iteration optimization when the weight $v_k$ of “update” item in Eq. \eqref{8} equals to 0. The total loss for the physics-based contrastive networks on $k^{th}$ client can be defined as follows:

\begin{equation}\mathcal{L}_k=\mathcal{L}^{r e c}+\frac{\tau}{2}\left\|\Theta_g-\Theta_k\right\|^2,\label{7}\end{equation}
where $\tau$ is applied to balance the weight parameters of reconstruction loss and regulation term.

In our SSFedMRI, the base network adopts MoDL [35], which can learn CNN-based priors to remove alias artifacts and noise. Since only the denoiser $D_\Theta$ of MR image $\mathrm{x}$ in Eq.\eqref{3a} contains learned parameters $\left
(\Theta\right)$, we train and transmit $\Theta$ between the server and clients.

\subsubsection{Soft updates for personalized model}

In classic FL, the trained local models $\Theta_k^{r}$ from all clients are aggregated via Eq.\eqref{5} to get the global model $\Theta_g^{r}$. The global model $\Theta_g^{r}$ in communication round ${r}$ is then transmitted back to local clients and used to initialize the local models $\Theta_k^{r+1}$ in communication round $r+1$. However, under the setting of non-i.i.d data (data acquired with different imaging protocols), domain shift will occur and result in performance degradation for local models. To mitigate performance degradation, we develop soft updates operation to calculate the weights of personalized denoiser models on the client side during each communication round. Specifically, we introduce an update item $\left(\Theta_g^r-\Theta_k^r\right)$ to measure the discrepancy between the global mode and each local model. The initialization of the personalized model on $k^{th}$ client in communication round ${r+1}$ is:

\begin{equation}\widehat{\Theta}_k^{r+1}=\Theta_k^r+v_k^r\left(\Theta_g^r-\Theta_k^r\right),\label{8}\end{equation}
where $v_k^r \in[0,1]$ is the weight parameter to balance the relationship between local model and the update item. $\Theta_g^r$ is the global model in communication round $r$, and $\Theta_k^r$ is local model in communication round $r$. When $v_k=0$, it means that the update of each local model only relies on the previous local model. When $v_k=1$, each local model will be identical to the global model, Eq. \eqref{8} reduces to the traditional FedAvg \cite{mcmahan2017communication}.

$v_k=1-\min \left(1, \sigma_k\left\|\Theta_g^r-\Theta_k^r\right\|\right)$, which changes according to the discrepancy between the global model and the local model. When $\left\|\Theta_g^r-\Theta_k^r\right\|$ is small, the update of the personalized model is influenced by the update item. When $\sigma_k\left\|\Theta_g^r-\Theta_k^r\right\|$ is large and $\sigma_k\left\|\Theta_g^r-\Theta_k^r\right\|>1$, the update of the personalized model is only determined by the local model. $\sigma_k$ is a weight to normalize the discrepancy:

\begin{equation}\sigma_k=\frac{\beta}{\left\|\Theta_g^r-\Theta_k^r\right\|},\label{9}\end{equation}
where $\beta$ is set to 0.8. For each client, $\sigma_k$ calculated at the first communication round $\left(r=1\right)$ and will be used throughout the experiments.

\section{Experiments}
\subsection{Experimental Setup}
\subsubsection{Datasets}
Four complex-value datasets, FastMRI \cite{knoll2020fastmri}, CC359\footnote{https://sites.google.com/view/calgary-campinas-dataset/mr-reconstructionchallenge}, Modl dataset\footnote{https://github.com/hkaggarwal/modl}, and one in-house dataset with different distributions are employed as local clients in our task of FL for MRI reconstruction. We randomly selected 10 $\%$ subjects from training, validation, and testing sets in FastMRI to run our experiments. The matrix size is 256 × 256.
\begin{itemize}
    \item FastMRI (F): T1-weighted brain MR images of 3526 subjects were adopted. Among the 3526 subjects, 2583 subjects were used for training, 83 subjects for validation, and 860 subjects for testing. Each subject contains about 16 image slices. Test dataset of T2-weighted brain MR images with 95 subjects were adopted, which were set as unseen data distribution.
    \item CC359 (C): T1-weighted brain MRI data of 35 subjects were adopted. Among these 35 subjects, 25 were used for training, 4 for validation, and 7 for testing. These data were acquired on a clinical MR scanner (Discovery MR750; General Electric (GE) Healthcare, Waukesha, WI) with a 12-channel imaging coil.
    \item Modl dataset \cite{aggarwal2018modl} (M): This dataset contains the fully sample brain MRI data of 5 subjects acquired using a T2 CUBE sequence. Among the 5 subjects, 3 subjects were used for training, 1 for validation, and 1 for testing.
    \item In-house dataset (I): Brain MR images of 100 subjects were collected. Among these 100 subjects, 80 were used for training, 8 for validation, and 12 for testing.
\end{itemize}

\subsubsection{Implementation details}
 Our SSFedMRI was implemented with the following hyper-parameters: communication rounds $R=200$, local epoch $L=4$ at first communication round, and local epoch $L=1$ at other communication rounds. The physics-based contrative reconstruction networks adopted the standard MoDL model \cite{aggarwal2018modl} with the number of iterations set to 10. Following the previous experience \cite{hu2021self}, the weight ($\gamma$ in Eq. \eqref{6}) for contrastive consistency loss was set to 0.01. The learning rate was 0.0001, and the models were optimized with RMSProp optimizer. The batch size was set to 4. All experiments were performed on four NVIDIA TITAN Xp GPUs (with 12GB memory). All training datasets and test datasets were normalized to the range of $[0,1]$, and the images were cropped to 256 × 256.
\subsubsection{Evaluation}
The peak signal-to-noise ratio (PSNR) and the structural similarity index (SSIM) between fully sampled reference images and reconstruction images are calculated to evaluate the quality of reconstruction images. Both quantitative results on individual clients and results averaged over the four clients are reported in the tables.
\subsubsection{Baselines}
We have compared the proposed SSFedMRI with state-of-the-arts to demonstrate the effectiveness of our method, including federated learning approaches and local self-supervised learning approaches. The federated approaches include: (1) FedAvg-SS \cite{mcmahan2017communication}, self-supervised learning without fully sampled data using the FedAvg strategy that averages the parameters of all local models to obtain a global model; (2) FedProx-SS \cite{li2020federated}, self-supervised learning without fully sampled data by aggregating local models with the FedProx strategy, which adds a proximal term to FedAvg; (3) FL-MRCM \cite{guo2021multi}, supervised learning with fully sampled data by aligning latent features between source clients and target domain to reduce the domain shift of non-i.i.d data; (4) FedMRI \cite{feng2022specificity}, supervised learning with fully sampled data by only aggregating the encoder network (not the decoder network) on local clients via the model contrastive FL strategy, which pulls the local model to the global model as well as pushes the local model at the current round away from that at the last round. In addition, the local self-supervised learning approaches include: (5) SSDU \cite{yaman2020self}, re-undersampling the acquired undersampled data to two subsets, one subset is used as input, and the other is treated as supervisory signals; (6) PARCEL \cite{wang2022parcel}, applying contrastive representation learning for MR imaging.

\section{Results}
\subsection{Generalizability evaluation on different clients}

To evaluate the model’s generalizability on different clients, we conduct experiments under four scenarios to test the model's capability in handling in-distribution and out-of-distribution datasets. In scenario 1 (Table \ref{tab1} Single), we train the model with data from a single client and evaluate its performance on the same client. In scenario 2 (Table \ref{tab1} Crossed), we train the model with data from a single client and evaluate its performance on another client. In scenario 3 (Table \ref{tab1} Federated), we collaboratively train the proposed SSFedMRI, and respectively evaluate the trained model on each client. In scenario 4 (Table \ref{tab2}), we extend the scenario 1 and 3 to evaluate performance of the trained model on unseen data distribution. In scenario 1-4, we employ two state-of-the-arts self-supervised learning approaches for MRI reconstruction, including SSDU \cite{yaman2020self} and PARCEL \cite{wang2022parcel}, to perform experiments.

\begin{table}
\begin{center}
\caption{Quantitative reconstruction results of different self-supervised learning methods, including SSDU, PARCEL and our proposed SSFedMRI, in which the models are trained by different strategies in scenario 1-3. Bold, underlined and starred values indicate the three best results, respectively.}
\label{tab1}
\begin{tabular}{c c c c c c}
\hline
Scenarios & Methods & Train & Test & PSNR & SSIM\\[0.3ex]
\hline
\multirow{8}{*}{Single}& \multirow{4}{*}{SSDU \cite{yaman2020self}} & F & F & 36.6542* & \underline{0.9778}\\[0.3ex]

\multicolumn{1}{c}{}&\multicolumn{1}{c}{}& C & C & 35.2624* & 0.9537*\\[0.3ex]
 
\multicolumn{1}{c}{}&\multicolumn{1}{c}{}& M & M & 22.6840* & 0.7984*\\[0.3ex]

\multicolumn{1}{c}{}&\multicolumn{1}{c}{}& I & I & 30.6867* & 0.8735*\\[0.3ex]

\cline{2-6} 
\multicolumn{1}{c}{}& \multirow{4}{*}{PARCEL \cite{wang2022parcel}} & F & F & \underline{37.5939} & 0.9684\\[0.3ex]

\multicolumn{1}{c}{}&\multicolumn{1}{c}{}& C & C & \underline{35.7521} & \underline{0.9541}\\[0.3ex]
 
\multicolumn{1}{c}{}&\multicolumn{1}{c}{}& M & M & \underline{26.9067} & \underline{0.8652}\\[0.3ex]

\multicolumn{1}{c}{}&\multicolumn{1}{c}{}& I & I & \underline{31.1508} & \underline{0.8800}\\[0.3ex]
 
\hline
\multirow{24}{*}{Crossed} & \multirow{12}{*}{SSDU} & C & F & 35.0044 & 0.9539\\[0.3ex]

\multicolumn{1}{c}{}&\multicolumn{1}{c}{}& M & F & 31.6922 & 0.9313\\[0.3ex]
 
\multicolumn{1}{c}{}&\multicolumn{1}{c}{}& I & F & 25.6199 & 0.9073\\[0.3ex]
 
\cline{3-6} 
\multicolumn{1}{c}{}&\multicolumn{1}{c}{}& F & C & 33.6373 & 0.9423\\[0.3ex]

\multicolumn{1}{c}{}&\multicolumn{1}{c}{}& M & C & 30.8496 & 0.8727\\[0.3ex]

\multicolumn{1}{c}{}&\multicolumn{1}{c}{}& I & C & 33.5346 & 0.9231\\[0.3ex]
 
\cline{3-6} 
\multicolumn{1}{c}{}&\multicolumn{1}{c}{}& F & M & 8.7446 & 0.4243\\[0.3ex]

\multicolumn{1}{c}{}&\multicolumn{1}{c}{}& C & M & 13.8333 & 0.5515\\[0.3ex]

\multicolumn{1}{c}{}&\multicolumn{1}{c}{}& I & M & 12.2027 & 0.6861\\[0.3ex]

\cline{3-6} 
\multicolumn{1}{c}{}&\multicolumn{1}{c}{}& F & I & 15.4952 & 0.5058\\[0.3ex]

\multicolumn{1}{c}{}&\multicolumn{1}{c}{}& C & I & 26.9075 & 0.7520\\[0.3ex]

\multicolumn{1}{c}{}&\multicolumn{1}{c}{}& M & I & 29.0970 & 0.8248\\[0.3ex]
 
\cline{2-6} 
\multicolumn{1}{c}{}& \multirow{12}{*}{PARCEL} & C & F & 34.6142 & 0.9471\\[0.3ex]

\multicolumn{1}{c}{}&\multicolumn{1}{c}{}& M & F & 34.6817 & 0.9519\\[0.3ex]
 
\multicolumn{1}{c}{}&\multicolumn{1}{c}{}& I & F & 25.7361 & 0.9109\\[0.3ex]
 
\cline{3-6} 
\multicolumn{1}{c}{}&\multicolumn{1}{c}{}& F & C & 33.5001 & 0.9173\\[0.3ex]

\multicolumn{1}{c}{}&\multicolumn{1}{c}{}& M & C & 32.2904 & 0.8986\\[0.3ex]

\multicolumn{1}{c}{}&\multicolumn{1}{c}{}& I & C & 34.1221 & 0.9289\\[0.3ex]
 
\cline{3-6} 
\multicolumn{1}{c}{}&\multicolumn{1}{c}{}& F & M & 18.1283 & 0.4941\\[0.3ex]

\multicolumn{1}{c}{}&\multicolumn{1}{c}{}& C & M & 12.2297 & 0.4418\\[0.3ex]

\multicolumn{1}{c}{}&\multicolumn{1}{c}{}& I & M & 9.1437 & 0.6705\\[0.3ex]

\cline{3-6} 
\multicolumn{1}{c}{}&\multicolumn{1}{c}{}& F & I & 19.9211 & 0.5169\\[0.3ex]

\multicolumn{1}{c}{}&\multicolumn{1}{c}{}& C & I & 18.7454 & 0.5611\\[0.3ex]

\multicolumn{1}{c}{}&\multicolumn{1}{c}{}& M & I & 29.8474 & 0.8452\\[0.3ex]

\hline
\multirow{4}{*}{Federated} & \multirow{4}{*}{SSFedMRI} & Fed & F & \textbf{41.1234} & \textbf{0.9811}\\[0.3ex]

\multicolumn{1}{c}{}&\multicolumn{1}{c}{}& Fed & C & \textbf{35.7687} & \textbf{0.9558}\\[0.3ex]
 
\multicolumn{1}{c}{}&\multicolumn{1}{c}{}& Fed & M & \textbf{33.3523} & \textbf{0.9106}\\[0.3ex]
\multicolumn{1}{c}{}&\multicolumn{1}{c}{}& Fed & I & \textbf{32.2695} & \textbf{0.8893}\\[0.3ex]
\hline
\end{tabular}
\end{center}
\end{table}

\begin{figure}[htbp] 
\centering
\includegraphics[width=0.5\textwidth]{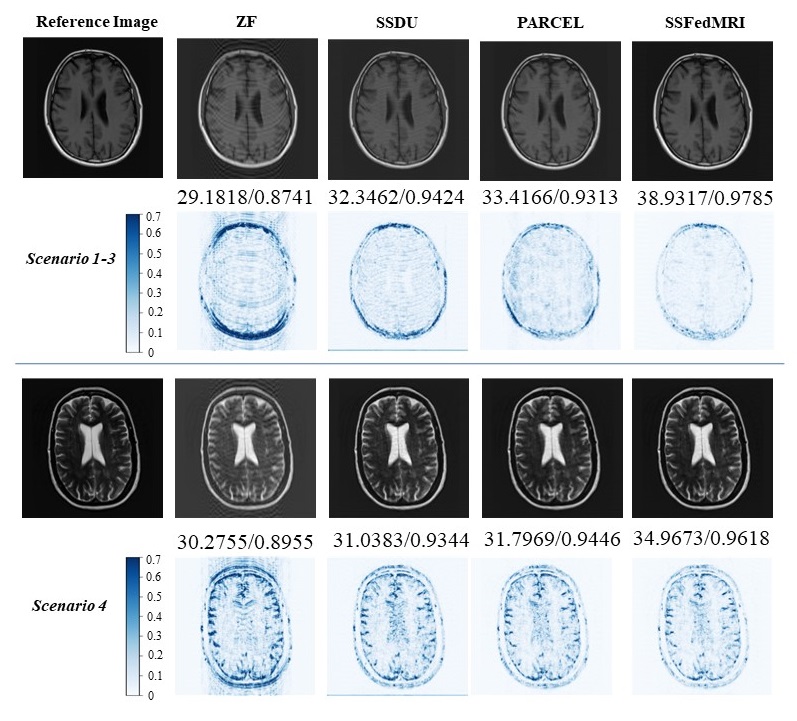} 
\caption{Reconstruction images and corresponding error maps of state-of-the-arts self-supervised learning methods (SSDU and PARCEL) and our proposed SSFedMRI. The upper half and bottom half show reconstruction results in scenario 1-3 and scenario 4 respectively.}
\label{fig2}
\end{figure}

Table \ref{tab1} lists the quantitative reconstruction results of different methods in scenario 1-3. The input undersampled data is acquired via 1D random sampling at 4-fold acceleration. It can be observed in Table \ref{tab1} that PARCEL under the setting of \textbf{Single} achieves better performances than that under the setting of \textbf{Crossed}. These results state that data distribution heterogeneity is existing in these four clients (corresponding to four datasets), and isolated training of PARCEL exhibits weaker generalization on other clients. In the setting of data heterogeneity, our proposed SSFedMRI exhibits better generalization on four participant clients than other methods in scenario 1 and 2. The upper half part of Fig. \ref{fig2} plots the three best reconstruction images in FastMRI as well as the corresponding error maps (with a display range of [0, 0.7]), which are indicated as bold, underlined and starred values in Table \ref{tab1}. It can be observed that our SSFedMRI achieves superior reconstruction results than SSDU and PARCEL. In the error maps, SSFedMRI presents fewer artifacts than other approaches, especially in edges of brain.

Table \ref{tab2} lists the quantitative reconstruction results of different methods in scenario 4, which uses the test dataset of T2-weighted images in FastMRI as our unseen test dataset, since it presents different data distribution. It can be observed that our proposed SSFedMRI outperforms other approaches trained with data from a single client, and SSFedMRI exhibits the best generalization on unseen data distribution. The bottom half part of Fig. \ref{fig2} plots the three best reconstruction images and error maps in test dataset, which are also indicated as bold, underlined and starred values in Table \ref{tab2}.

\begin{table}
\begin{center}
\caption{Quantitative reconstruction results of different self-supervised learning methods, including SSDU, PARCEL and our proposed SSFedMRI, in which the trained models are evaluated on unseen data distribution under scenario 4. Bold, underlined and starred values indicate the three best results, respectively.}
\label{tab2}
\begin{tabular}{c c c c c}
\hline
Methods & Train & Test & PSNR & SSIM\\[0.3ex]
\hline
\multirow{4}{*}{SSDU} & F & Unseen & 32.7881* & \underline{0.9603}\\[0.3ex]

\multicolumn{1}{c}{}& C & Unseen & 30.5107 & 0.9516\\[0.3ex]
 
\multicolumn{1}{c}{}& M & Unseen & 29.4158 & 0.9147\\[0.3ex]

\multicolumn{1}{c}{}& I & Unseen & 22.3749 & 0.8914\\[0.3ex]

\hline
\multirow{4}{*}{PARCEL} & F & Unseen & \underline{33.2103} & 0.9608*\\[0.3ex]

\multicolumn{1}{c}{}& C & Unseen & 31.8522 & 0.9230\\[0.3ex]
 
\multicolumn{1}{c}{}& M & Unseen & 31.7672 & 0.9314\\[0.3ex]

\multicolumn{1}{c}{}& I & Unseen & 23.9821 & 0.9022\\[0.3ex]

\hline
SSFedMRI & Fed & Unseen & \textbf{36.0824} & \textbf{0.9615}\\[0.3ex]
\hline
\end{tabular}
\end{center}
\end{table}

\subsection{Comparison with supervised FL methods}
\subsubsection{Reconstruction accuracy}

\begin{figure}[htbp] 
\centering
\includegraphics[width=0.5\textwidth]{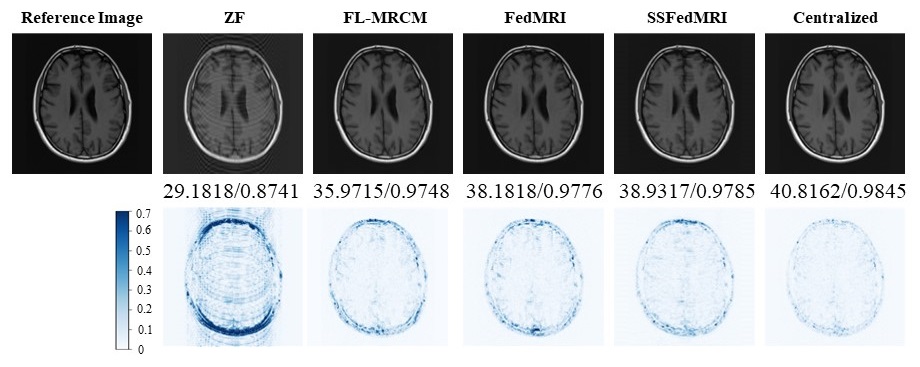} 
\caption{Reconstruction images and corresponding error maps of different methods (FL-MRCM, FedMRI, proposed SSFedMRI and \textit{Centralized}) at 4-fold acceleration (1D random undersampling).}
\label{fig3}
\end{figure}

In this section, we evaluate the reconstruction accuracy among various FL methods and centralized training, where FL-MRCM and FedMRI are existing FL-based methods for MRI reconstruction, and \textit{Centralized} indicates that all local data in participant clients are gathered together. \textit{Centralized} is regarded as upper bound of FL methods, in which the model is trained via supervised learning. To comprehensively compare reconstruction accuracy of these methods with different amounts of training data, we perform experiments under two scenarios. In scenario 5, the reconstruction models are trained with the number of local training dataset described in section IV (4128 slices, 2750 slices, 360 slices, and 2634 slices for training, respectively). In scenario 6, the local training datasets are further reduced to 400 slices, 220 slices, 360 slices and 450 slices, respectively.
The first sub-table of Table \ref{tab3} lists the quantitative results (PSNR and SSIM) of different approaches under scenario 5. Similarly, the average values of PSNR and SSIM achieved by our proposed SSFedMRI are comparable with or slightly higher than the two existing supervised FL-based methods for MRI reconstruction (FL-MRCM and FedMRI). Fig. \ref{fig3} plots the visual reconstruction results and the corresponding error maps (with a display range of [0, 0.7]) of above methods under scenario 5. From the qualitative results, we can observe that our proposed SSFedMRI, which trains in a self-supervised manner, can achieve comparable reconstruction performance when compared to the two supervised federated learning methods. The error maps also reveal that SSFedMRI achieves competitive reconstruction results, especially on the edge of the brain image, the edge reconstructed using SSFedMRI are clearer than that of the others. 

\begin{table*}[ht]
\begin{center}
\caption{Quantitative reconstruction results of different methods, including state-of-the-art supervised federated learning methods (FL-MRCM and FedMRI), our proposed SSFedMRI and \textit{Centralized} under scenario 5 and 6. The input is undersampled MRI data with 1D random mask at 4-fold acceleration.}
\label{tab3}
\begin{tabular}{c c c c c c c c c c c }
\hline
\multicolumn{11}{c}{Scenario 5}\\[0.3ex]
\hline
\multirow{3}{*}{Methods} & \multicolumn{2}{c}{F} & \multicolumn{2}{c}{C} & \multicolumn{2}{c}{M} & \multicolumn{2}{c}{I} & \multicolumn{2}{c}{\multirow{2}{*}{Avg.}}\\[0.3ex]

\multicolumn{1}{c}{}& \multicolumn{2}{c}{4128 slices} & \multicolumn{2}{c}{2750 slices} & \multicolumn{2}{c}{360 slices} & \multicolumn{2}{c}{2634 slices} & \multicolumn{1}{c}{}\\[0.3ex]
\cline{2-11}
\multicolumn{1}{c}{}& PSNR & SSIM & PSNR & SSIM & PSNR & SSIM & PSNR & SSIM & PSNR & SSIM\\[0.3ex]
\hline
Centralized (Ub) & 43.7950 & 0.9881 & 37.3909 & 0.9621 & 43.4279 & 0.9776 & 32.3651 & 0.8944 & 39.2447 & 0.9555\\[0.3ex]
\hline
FL-MRCM \cite{guo2021multi} & 36.7675 & 0.9676 & 33.2196 & 0.9329 & 29.5928 & 0.8356 & 30.0720 & 0.8775 & 32.4129 & 0.9034\\[0.3ex]

FedMRI \cite{feng2022specificity} & 38.2359 & 0.9711 & 32.6401 & 0.9272 & 30.2502 & 0.8264 & 30.1891 & 0.8826 & 32.8288 & 0.9018\\[0.3ex]

SSFedMRI & \textbf{41.1234} & \textbf{0.9811} & \textbf{35.7687} & \textbf{0.9558} & \textbf{33.3523} & \textbf{0.9106} & \textbf{32.2695} & \textbf{0.8893} & \textbf{35.6285} & \textbf{0.9342}\\[0.3ex]
\hline
\multicolumn{11}{c}{Scenario 6}\\[0.3ex]
\hline
\multirow{3}{*}{Methods} & \multicolumn{2}{c}{F} & \multicolumn{2}{c}{C} & \multicolumn{2}{c}{M} & \multicolumn{2}{c}{I} & \multicolumn{2}{c}{\multirow{2}{*}{Avg.}}\\[0.3ex]

\multicolumn{1}{c}{}& \multicolumn{2}{c}{400 slices} & \multicolumn{2}{c}{220 slices} & \multicolumn{2}{c}{360 slices} & \multicolumn{2}{c}{450 slices} & \multicolumn{1}{c}{}\\[0.3ex]
\cline{2-11}
\multicolumn{1}{c}{}& PSNR & SSIM & PSNR & SSIM & PSNR & SSIM & PSNR & SSIM & PSNR & SSIM\\[0.3ex]

\hline
Centralized (Ub) & 43.1014 & 0.9874 & 37.1647 & 0.9619 & 43.3374 & 0.9775 & 31.3701 & 0.8778 & 38.7434 & 0.9511\\[0.3ex]
\hline
FL-MRCM & 30.6721 & 0.9392 & 28.1240 & 0.8795 & 27.9137 & 0.8025 & 25.9134 & 0.7461 & 28.1558 & 0.8418\\[0.3ex]

FedMRI & 36.5408 & 0.9619 & 30.8655 & 0.9031 & 30.1436 & 0.8248 & 28.6709 & 0.8391 & 31.5552 & 0.8822\\[0.3ex]

SSFedMRI & \textbf{39.5992} & \textbf{0.9745} & \textbf{35.0717} & \textbf{0.9530} & \textbf{33.6161} & \textbf{0.9154} & \textbf{30.0761} & \textbf{0.8376} & \textbf{34.5908} & \textbf{0.9201}\\[0.3ex]
\hline
\end{tabular}
\end{center}
\end{table*}

Quantitative results in scenario 6 are listed in the second sub-table of Table \ref{tab3}. Under this more challenging experimental setting, our SSFedMRI performs the best when compared to FL-MRCM and FedMRI. In addition, the performance enhancement brought by SSFedMRI is larger than the setting under scenario 5. Specifically, in scenario 5, the average value of PSNR achieved by SSFedMRI is 9.92$\%$ higher than that of FL-MRCM. In scenario 6, the average value of PSNR achieved by SSFedMRI is 22.85$\%$ higher than that of FL-MRCM. Therefore, we conclude that SSFedMRI is more effective when limited local training data are provided. The reason for above results is that the FL-MRCM and FedMRI employs U-Net as base network, their reconstruction performances rely on sufficient local training data. But in SSFedMRI, MoDL \cite{aggarwal2018modl} is employed as base network, which exhibits more superiority when the training data is reduced due to the application of MRI’s physics properties.

\subsubsection{Communication cost}

In classic FL (e.g., FedAvg), each client uploads their model parameters to the server, and then, downloads the aggregated model parameters from the server to update the client’s model in each communication round. The size of messages transmitted though the clients and the server can be expressed as

\begin{equation}cost=2 K|w| T,\label{10}\end{equation}
which includes the upload and download of participant clients. $K$ is the number of participant clients, $|w|$ is the parameter size of network in each client, and $T$ denotes the training round.

Compared with the existing FL approaches for MRI reconstruction, such as FL-MRCM and FedMRI, which adopt U-Net as the base network in clients, $|w|$ in SSFedMRI is much smaller (the base network is MoDL). Table \ref{tab4} shows the network parameter counts of different methods. We can observe that the network parameters of global model in SSFedMRI is about 2.96$\%$ of that in FL-MRCM, and about 4.88$\%$ of that in FedMRI. Compared to the supervised learning counterparts, the communication cost of SSFedMRI is still the lowest, which is helpful when the transmission bandwidth is limited.

\begin{table}[ht]
\begin{center}
\caption{Network parameters of different methods.}
\label{tab4}
\begin{tabular}{c c c}
\hline
Methods & \multicolumn{2}{c}{Parameters}\\[0.3ex]

  & Client & Server\\[0.3ex]
\hline
FL-MRCM & 7.76 M & 7.76 M\\[0.3ex]

FedMRI & 7.76 M & 4.71 M\\[0.3ex]

SSFedMRI & \textbf{0.23 M} & \textbf{0.23 M}\\[0.3ex]
\hline
\end{tabular}
\end{center}
\end{table}

\subsection{Ablation study}

\begin{table*}[ht]
\begin{center}
\caption{Quantitative reconstruction results of different self-supervised federated learning methods, including Fedprox SS, FedAvg SS and our proposed SSFedMRI. The input is undersampled with 1D random mask for 4-fold acceleration.}
\label{tab5}
\begin{tabular}{c c c c c c c c c c c }
\hline
\multirow{2}{*}{Methods} & \multicolumn{2}{c}{F} & \multicolumn{2}{c}{C} & \multicolumn{2}{c}{M} & \multicolumn{2}{c}{I} & \multicolumn{2}{c}{Avg.}\\[0.3ex]

\multicolumn{1}{c}{}& PSNR & SSIM & PSNR & SSIM & PSNR & SSIM & PSNR & SSIM & PSNR & SSIM\\[0.3ex]
\hline
FedProx SS \cite{li2020federated} & 39.5964 & 0.9748 & 34.2022 & 0.9441 & 32.1224 & 0.8881 & 30.0443 & 0.8361 & 33.9913 & 0.9108\\[0.3ex]

FedAvg SS \cite{mcmahan2017communication} & 40.5806 & 0.9787 & 35.3170 & 0.9529 & 32.7427 & 0.9011 & 31.4772 & 0.8763 & 35.0294 & 0.9272\\[0.3ex]

SSFedMRI & \textbf{41.1234} & \textbf{0.9811} & \textbf{35.7687} & \textbf{0.9558} & \textbf{33.3523} & \textbf{0.9106} & \textbf{32.2695} & \textbf{0.8893} & \textbf{35.6285} & \textbf{0.9342}\\[0.3ex]
\hline
\end{tabular}
\end{center}
\end{table*}

\begin{figure}[htbp]
\centering
\includegraphics[width=0.5\textwidth]{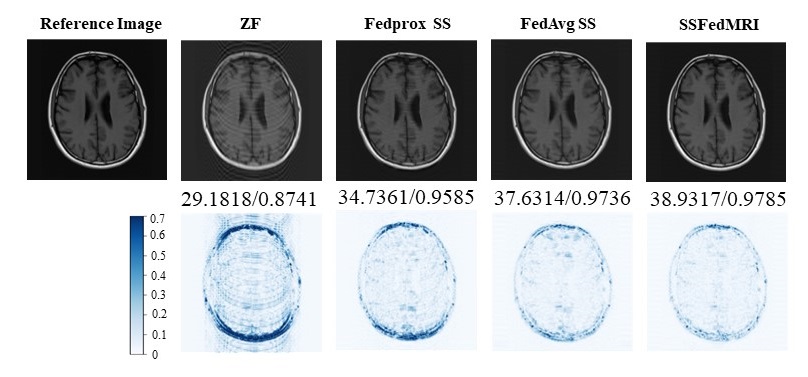} 
\caption{Reconstruction images and corresponding error maps of different self-supervised federated learning methods, including FedProx SS, FedAvg SS and our proposed SSFedMRI.}
\label{fig4}
\end{figure}

In this section, we conduct experiments to assess the effectiveness of personalized soft updates in SSFedMRI. First, we provide the results of SSFedMRI. Then, we remove the personalized soft update operation (FedProx SS) from SSFedMRI. Here, the global model is applied to overwrite the local models before local training in each communication round. Finally, we remove both the personalized soft update operation and the regularization item in Eq. \eqref{7}, which actually degrades into classic federated learning with self-supervised learning (FedAvg SS). Fig. \ref{fig4} and Table \ref{tab5} give the qualitative and quantitative results of the above variants, respectively. The undersampled data is also acquired via 1D random sampling at 4-fold acceleration. According to the quantitative results (Table \ref{tab5}), the average values of PSNR and SSIM achieved by SSFedMRI are the highest among the three variants, and on each client, the PSNR and SSIM of SSFedMRI are also higher than other variants. Furthermore, we observe that the performances of Fedprox SS are worse than that of FedAvg SS, which may be affected by imbalance data quality and different capabilities of local models by self-supervised learning. Even so, the PSNR of SSFedMRI have increased 4.8$\%$ and 1.7$\%$ over that of FedProx SS and FedAvg SS respectively. The improvements of SSFedMRI are due to the contribution of soft update operation. From the results in Fig. \ref{fig4} and Table \ref{tab5}, we confirm that although FL is crucial for learning with multi-institutional data, much consideration should be paid to the design of the communication between the global model and the local models.

\section{Conclusion}
In this work, we aim to solve the issue of fully sampled data deficiency in FL for MRI reconstruction. We propose a self-supervised federated learning method, SSFedMRI. Physics-based contrastive reconstruction networks are built on each client to achieve cross-site collaborative training in the absence of fully sampled reference data without sharing data. Compared with existing FL-based methods, reduced communication cost in FL is achieved due to the decreased number of network parameters. To further mitigate performance degradation caused by data heterogeneity in FL, a personalized soft update operation is proposed to softly update local models. Particularly, in each communication round, instead of utilizing the global model to overwrite the local models, SSFedMRI enforces local models to consider the discrepancies between the local models and the global model, forcing local models to learn cross-site prior information from participating clients while still holding optimization towards the local data distributions. Extensive experiments on multi-site complex-valued MRI datasets acquired with different imaging protocols have been performed. Both qualitative and quantitative results demonstrate that our SSFedMRI can achieve better performances.



\bibliographystyle{IEEEtran}
\bibliography{IEEEabrv,main}

\end{document}